\documentclass[aps,prl,twocolumn,superscriptaddress,floatfix,nofootinbib,showpacs,amsmath,amssymb,altaffilletter]{revtex4-1}

\usepackage{graphicx}
\usepackage{dcolumn}
\usepackage{bm}
\usepackage{color}
\usepackage{verbatim}
\usepackage{paralist}
\usepackage{lineno}
\usepackage{subfigure}

\begin{document}
\title{Spectroscopy of geo-neutrinos from 2056 days of Borexino data}

\newcommand{\APC}{Laboratoire AstroParticule et Cosmologie, 75231 Paris cedex 13, France}
\newcommand{\Dubna}{Joint Institute for Nuclear Research, 141980 Dubna, Russia}
\newcommand{\Genova}{Dipartimento di Fisica, Universit\`a degli Studi e INFN, Genova 16146, Italy}
\newcommand{\Hamburg}{Institut f\"ur Experimentalphysik, Universit\"at, 22761 Hamburg, Germany}
\newcommand{\Heidelberg}{Max-Planck-Institut f\"ur Kernphysik, 69117 Heidelberg, Germany}
\newcommand{\Kiev}{Kiev Institute for Nuclear Research, 06380 Kiev, Ukraine}
\newcommand{\Krakow}{M.~Smoluchowski Institute of Physics, Jagiellonian University, 30059 Krakow, Poland}
\newcommand{\Kurchatov}{NRC Kurchatov Institute, 123182 Moscow, Russia}
\newcommand{\Kurchatovb}{ National Research Nuclear University MEPhI (Moscow Engineering Physics Institute), 115409 Moscow, Russia. }
\newcommand{\LNGS}{INFN Laboratori Nazionali del Gran Sasso, 67010 Assergi (AQ), Italy}
\newcommand{\Milano}{Dipartimento di Fisica, Universit\`a degli Studi e INFN, 20133 Milano, Italy}
\newcommand{\Perugia}{Dipartimento di Chimica, Universit\`a e INFN, 06123 Perugia, Italy}
\newcommand{\Peters}{St. Petersburg Nuclear Physics Institute NRC Kurchatov Institute, 188350 Gatchina, Russia}
\newcommand{\Princeton}{Physics Department, Princeton University, Princeton, NJ 08544, USA}
\newcommand{\PrincetonChemEng}{Chemical Engineering Department, Princeton University, Princeton, NJ 08544, USA}
\newcommand{\Queens}{Physics Department, Queen's University, Kingston ON K7L 3N6, Canada}
\newcommand{\UMass}{Amherst Center for Fundamental Interactions and Physics Department, University of Massachusetts, Amherst, MA 01003, USA}
\newcommand{\Virginia}{Physics Department, Virginia Polytechnic Institute and State University, Blacksburg, VA 24061, USA}
\newcommand{\Ferrara}{Dipartimento di Fisica e Scienze della Terra  Universit\`a degli Studi di Ferrara e INFN,  Via Saragat 1-44122, Ferrara, Ital }
\newcommand{\Munchen}{Physik-Department and Excellence Cluster Universe, Technische Universit\"at  M\"unchen, 85748 Garching, Germany. }
\newcommand{\Lomonosov}{ Lomonosov Moscow State University Skobeltsyn Institute of Nuclear Physics, 119234 Moscow, Russia. }
\newcommand{\GSSI}{ Gran Sasso Science Institute (INFN), 67100 \L'Aquila, Italy. }
\newcommand{\Huston}{Department of Physics, University of Houston, Houston, TX 77204, USA. }
\newcommand{\Dresda}{Department of Physics, Technische Universit\"at Dresden, 01062 Dresden, Germany. }
\newcommand{\UCLA}{Physics and Astronomy Department, University of California Los Angeles (UCLA), Los Angeles, California 90095, USA. }
\newcommand{\Honolulu}{Department of Physics and Astronomy, University of HawaiÕi, Honolulu, HI 96822, USA.}
\newcommand{\Mainz}{Institute of Physics and Excellence Cluster PRISMA, Johannes Gutenberg-Universit\"at Mainz, 55099 Mainz, Germany.}

\author{M.~Agostini}\affiliation{\Munchen}
\author{S.~Appel}\affiliation{\Munchen}
\author{G.~Bellini}\affiliation{\Milano}
\author{J.~Benziger}\affiliation{\PrincetonChemEng}
\author{D.~Bick}\affiliation{\Hamburg}
\author{G.~Bonfini}\affiliation{\LNGS}
\author{D.~Bravo}\affiliation{\Virginia}
\author{B.~Caccianiga}\affiliation{\Milano}
\author{F.~Calaprice}\affiliation{\Princeton}
\author{A.~Caminata}\affiliation{\Genova}
% ...

\collaboration{Borexino Collaboration}
\noaffiliation

\date{\today}

\begin{abstract}
We report an improved geo-neutrino measurement with Borexino from 2056 days of data taking. The present exposure is $(5.5\pm0.3)\times10^{31}$ proton$\times$yr. Assuming a chondritic Th/U mass ratio of 3.9, we obtain $23.7~^{+6.5}_{-5.7} (stat)~ ^{+0.9}_{-0.6} (sys)$ geo-neutrino events. The null observation of geo-neutrinos with Borexino alone has a probability of  $3.6 \times 10^{-9}$ (5.9$\sigma$).  A geo-neutrino signal from the mantle is obtained at 98\% C.L. The radiogenic heat production for U and Th from the present best-fit result is restricted to the range 23-36~TW, taking into account the uncertainty on the distribution of heat producing elements inside the Earth.
\end{abstract}

\keywords{Geo-neutrinos; Low background detectors; Liquid scintillators}
\pacs{13.35.Hb, 14.60.St, 26.65.+t, 95.55.Vj, 29.40.Mc}

\maketitle

Geo-neutrinos are electron anti-neutrinos ($\bar{\nu}_e$) produced by $\beta$ decays of long-lived isotopes, which are naturally present in the interior of the Earth,  such as decays in the $^{238}$U and $^{232}$Th chains, and $^{40}$K  \cite{georeview,georeview2}. 
Geo-$\bar{\nu}_e$ measurements have been reported by Borexino \cite{antinupaper1, antinupaper2} and KamLAND \cite{KamLANDpaper1, KamLANDpaper2}.
Here we present improved reactor neutrino and geo-$\bar{\nu}_e$ measurements performed by Borexino.  Borexino is an unsegmented liquid scintillator detector in operation at the underground Gran Sasso National Laboratory, Italy~\cite{bib:bxdetectorpaper}. 
In liquid scintillator detectors $\bar{\nu}_e$ are detected via the inverse $\beta$ decay process, $\bar{\nu}_e + p \rightarrow e^+ + n$, with a threshold of 1.806 MeV. 
The deposited $\bar{\nu}_e$ energy results in a prompt signal induced by the positron, and includes annihilation photons. The visible energy is related to the $\bar{\nu}_e$ energy  as  $E_{vis} = E_{\bar{\nu}_e} -  0.784$ MeV. 
A delayed signal induced by the neutron capture on protons produces the 2.22 MeV gamma-ray, providing a delayed coincidence signal with a mean capture time of 259.7$\pm$1.3(stat)$\pm$2.0(syst) $\mu$s \cite{neutrontau}.
In Borexino 278 tons of ultra-pure organic liquid scintillator (pseudocumene doped with 1.5 g/l of diphenyloxazole) is confined within a thin, spherical nylon inner vessel (IV) with a nominal radius of 4.25 m and concentrically placed inside a stainless steel sphere (SSS) with radius of 13.7 m and equipped with 2212 8" photomultipliers (PMTs). 
The volume between the SSS and the IV, divided in two regions by a second outer nylon vessel (OV) is filled with 890 tons of pseudocumene with 3 g/l of light quencher dimethylphthalate to shield the core of the detector against $\gamma$ radiation and radon from the PMTs and the SSS. 
The SSS is immersed in an water tank (WT) instrumented with 208 PMTs as a \v{C}erenkov active muon veto, also acting as an shield against $\gamma$-rays and and neutrons from the surrounding rock. 

In Borexino,  known sources of $\bar{\nu}_e$ events are nuclear reactors and geo-neutrinos. 
For geo-neutrinos, $^{238}$U and $^{232}$Th are the only isotopes abundant enough to significantly contribute events in Borexino above the inverse $\beta$ decay threshold. The light yield in Borexino is measured to be 
about 500 photoelectrons (p.e.) / MeV and the energy resolution scales as $\sim$5\%/$\sqrt(E)$ \cite{bib:bxdetectorpaper}. The Borexino liquid scintillator shows a high pulse shape discrimination efficiency in separating high ionizing particles (protons, $\alpha$'s) from electrons and gamma-rays \cite{longpaper}.

The data reported here were collected between 
December 15, 2007 and March 8, 2015 for a total of 2055.9 days before any selection cut. 
The number of PMTs for the present data set has been declining with time, from 1931 to 1525 (average 1730), with small run by run fluctuations.  
As reported in \cite{antinupaper1} the event energy is a calibrated non-linear function of the number of detected p.e. We perform the analysis directly in number of p.e. 
We discard events occurring within 2 ms of every muon crossing the outer detector and within 2 s of muons crossing the inner detector to reject neutrons and long-lived cosmogenic radioactivity, respectively. This cut reduces the live-time to 1841.9 days.
We apply the following additional selection cuts: 
(1) prompt scintillation light: $Q_p > $ 408 p.e. (i.e. 1.022 MeV corrected for the energy resolution); (2) delayed signal scintillation light: 860 $< Q_d <$ 1300 p.e. (neutron capture peak) ; (3) correlation distance between prompt and delayed signals: $\Delta R <$ 1m; (4) correlated time between prompt and delayed signals: 20 $< \Delta t < 1280 \, \mu s$; (5) pulse shape discrimination with Gatti filter \cite{gattipaper}: $g_{\alpha \beta} < 0.015$ for delayed signals; 
(6) multiplicity cut: selected event are neither preceded or followed by neutron-like events within a 2 ms window; (7) dynamical fiducial volume \cite{longpaper}: every prompt signal has a reconstructed vertex $>$30 cm away from the time-varying IV surface; (8) FADC cut: independent check of candidate events features by a 400 MHz digitizer acquisition system. The combined efficiency of the cuts is determined by Monte Carlo to be (84.2$\pm$1.5)\%. 
The total efficiency-corrected exposure for the present data set is 907$\pm$44 ton$\times$yr.
We have identified 77 $\bar{\nu}_e$ candidates passing all the selection cuts.

\begin{table}[h]
\begin{center}
\caption{Estimated backgrounds for $\bar{\nu}_e$ given in number of events. Upper limits are given for 90\% C.L.}
\begin{tabular}{lc}
\hline\hline
$^9$Li-$^8$He		          &0.194$^{+0.125}_{-0.089}$\\
Accidental coincidences	&0.221$\pm$0.004 \\
Time correlated		&0.035$^{+0.029}_{-0.028}$\\
($\alpha$,n) in scintillator		         &0.165$\pm$0.010\\
($\alpha$,n) in buffer	         &$<$0.51\\
Fast n's ($\mu$ in WT)                     & $<$0.01 \\
Fast n's ($\mu$ in rock)             & $<$ 0.43 \\
untagged muons                        & 0.12$\pm$0.01 \\
Fission in PMTs                          & 0.032$\pm$0.003 \\
$^{214}$Bi-$^{214}$Po       & 0.009$\pm$0.013 \\ \hline
Total                                         & 0.78$^{+0.13}_{-0.10}$ \\ 
					   & $<$ 0.65(combined) \\
\hline\hline
\end{tabular}
\label{tab:background}
\end{center}
\end{table}

The probability of $\bar{\nu}_e$-mimicking background events leaking into the dataset was evaluated as follows:
(1) The rate of accidental coincidences has been searched for by shifting the delayed time window to 2-20 seconds and keeping all other cuts unchanged. The energy spectrum of these events is limited to $<$3~MeV. 
(2) Time correlated events have been searched for in the (2 ms, 2 s) time window. A negligible amount of correlated events with a $\sim$1 s time constant were identified and their contribution in the $\bar{\nu}_e$ time window determined. 
(3) ($\alpha$,n) background for $\bar{\nu}_e$ search has been extensively discussed elsewhere \cite{antinupaper1, antinupaper2,KamLANDpaper1, KamLANDpaper2}. 
The average rate of  $^{210}$Po in the dataset is determined to be  (14.1$\pm$0.2) counts/(day$\cdot$ton).
(4) Cosmogenic radioactive isotopes which decay via $\beta$+n, namely $^9$Li-$^8$He \cite{neutrontau}, have been studied in the (2 ms, 2 s) time window after a muon crossing the inner detector. These events have a wide energy spectrum with maximum  at $\sim$5~MeV. 
(5) $^{222}$Rn in the liquid scintillator can produce background events through time-correlated $\beta + (\alpha + \gamma)$ decays of $^{214}$Bi and $^{214}$Po. This decay sequence has a time constant close to the neutron capture time following a $\bar{\nu}_e$ interaction. This background has been estimated by individually tagging  $^{214}$Bi-$^{214}$Po decays.
(6) Backgrounds from fast neutrons, untagged muons and spontaneous fission decays in the PMTs are the same as in previous papers \cite{antinupaper1, antinupaper2}.
Table \ref{tab:background} summarizes the estimated backgrounds for $\bar{\nu}_e$ candidates, expressed in number of events. The combined upper limit is obtained by Monte Carlo. The signal-to-background ratio is $\sim$100. 

\begin{figure}[!t]
\includegraphics[width=0.45\textwidth]{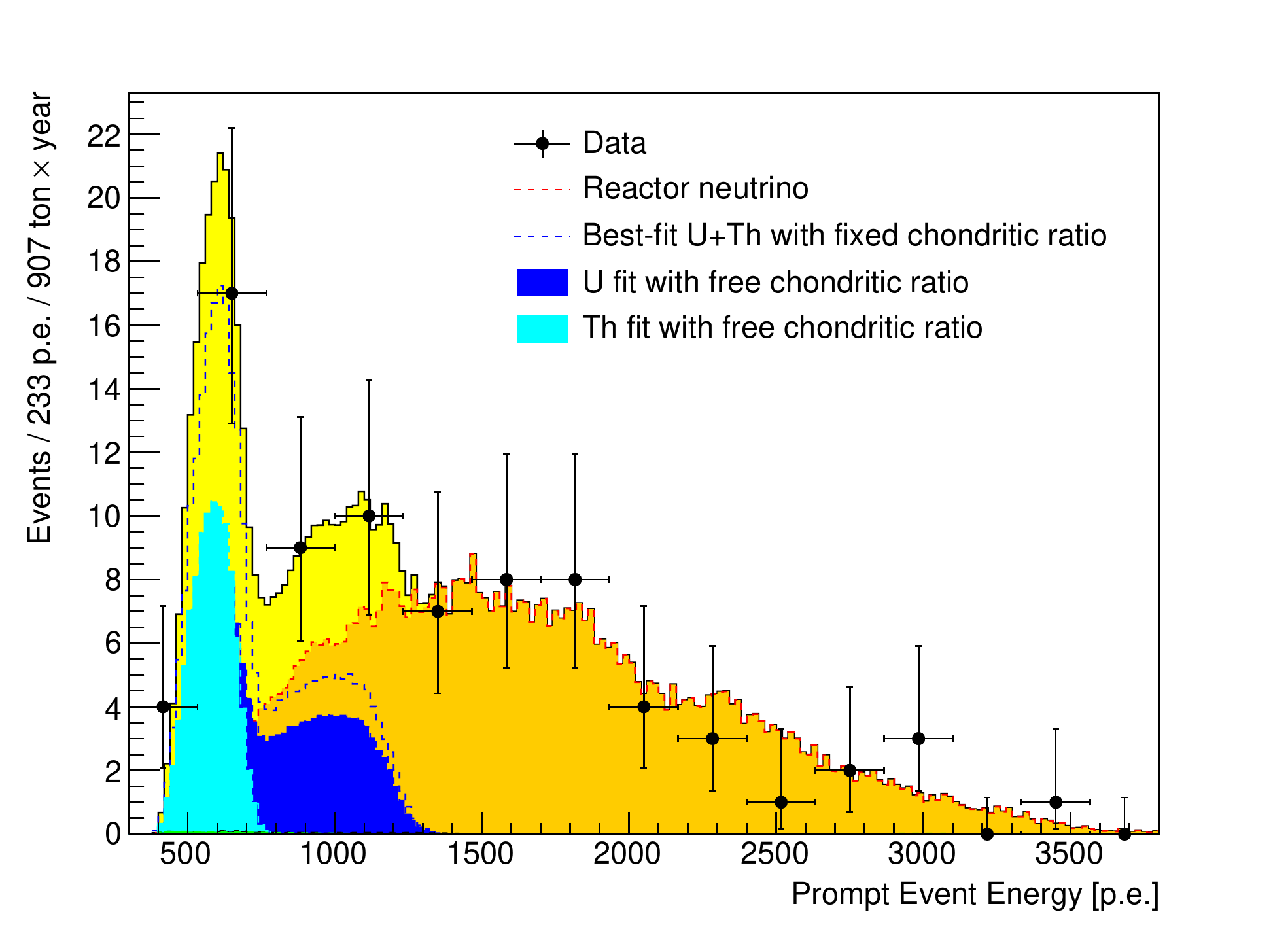}
\caption{Prompt light yield spectrum, in units of photoelectrons (p.e.), of $\bar{\nu}_e$ candidates and the best-fit. The best-fit shows the geo-neutrino and reactor neutrino spectra (dotted lines) assuming the chondritic ratio. Colored areas show the result of a separate fit with U (blue) and Th (light blue) set as free and independent parameters.}
\label{fig1}
\end{figure}

We have performed an un-binned likelihood fit of the energy spectrum of selected prompt $\bar{\nu}_e$ candidate events \cite{antinupaper1}, shown in Fig. \ref{fig1}.  
The reactor and geo-neutrinos spectra are obtained by Monte Carlo and the backgrounds considered in this analysis are reported in Table \ref{tab:background}. The Monte Carlo spectra have been determined as reported in \cite{antinupaper2}.
The reactor neutrinos signal has been calculated adopting the data from IAEA \cite{IAEA} updated to 2014 and the method described in \cite{Baldoncini}. For the first quarter of 2015 we have used the values from 2014. For the present exposure we predict ($87\pm 4$) TNU events from nuclear reactors, where 1 Terrestrial Neutrino Unit (TNU) = 1 event / year / 10$^{32}$ protons.
The log-likelihood function has two signal components, $S_{geo}$ and $S_{react}$, left free, and three background components, $S_{LiHe}$, $S_{\alpha n}$, $S_{acc}$, constrained to the values and errors reported in Tab.~\ref{tab:background}. These components account for 75\% of the total background. The other components were left out due to the uncertainty in their energy spectrum. Combined, they contribute $\sim$1\% to the best fit and their contribution to the systematic uncertainty is absorbed in the uncertainty on the energy scale. 

Using the value ratio for the masses of Th and U, $m(Th)/m(U) = 3.9$, suggested by the chondritic model, our best fit yields $S_{geo}=23.7^{+6.5}_{-5.7} (stat) ^{+0.9}_{-0.6} (sys) $ events $(43.5^{+11.8}_{-10.4} (stat) ^{+2.7}_{-2.4} (sys)$ TNU) and $S_{react}=52.7^{+8.5}_{-7.7}(stat) ^{+0.7}_{-0.9} (sys)$ events $(96.6^{+15.6}_{-14.2}(stat) ^{+4.9}_{-5.0} (sys)$ TNU). 
When expressing the results in TNU, systematic uncertainties from both the exposure (4.8\%) and the Monte Carlo energy calibration (1\%) are included.
Only the Monte Carlo calibration uncertainty is relevant when using the number of decays.

\begin{figure}[!t]
\includegraphics[width=0.4\textwidth]{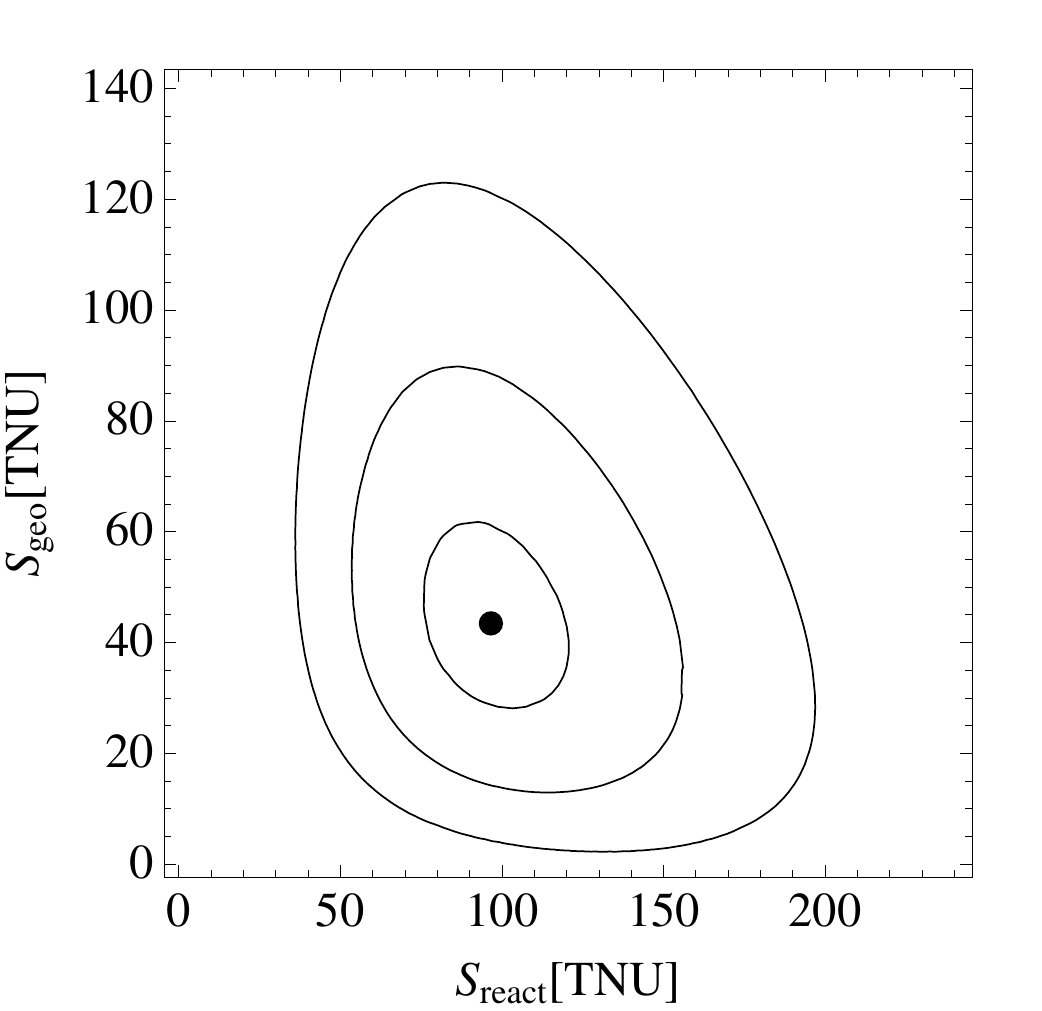}
\caption{Best-fit contours for 1, 3 and 5$\sigma$ for the statistics reported in this paper.} 
\label{fig2}
\end{figure}

In Fig. \ref{fig2} we show the 1, 3 and 5$\sigma$ contours from the log-likelihood fit.
Borexino alone observes geo-neutrinos with 5.9$\sigma$ significance (Fig. \ref{fig2}). The null hypothesis for geo-neutrino observation has a probability equal to $3.6 \times 10^{-9}$.
The measured geo-neutrino signal corresponds to  $\bar{\nu}_e$ fluxes at the detector from decays in the U and Th chains of $\phi(U)=(2.7\pm0.7)\times10^6$ cm$^{-2}$s$^{-1}$ and $\phi(Th)=(2.3\pm0.6)\times10^6$ cm$^{-2}$s$^{-1}$, respectively. Statistical and systematic uncertainties are added in quadrature.

\begin{figure}[!t]
\includegraphics[width=0.4\textwidth]{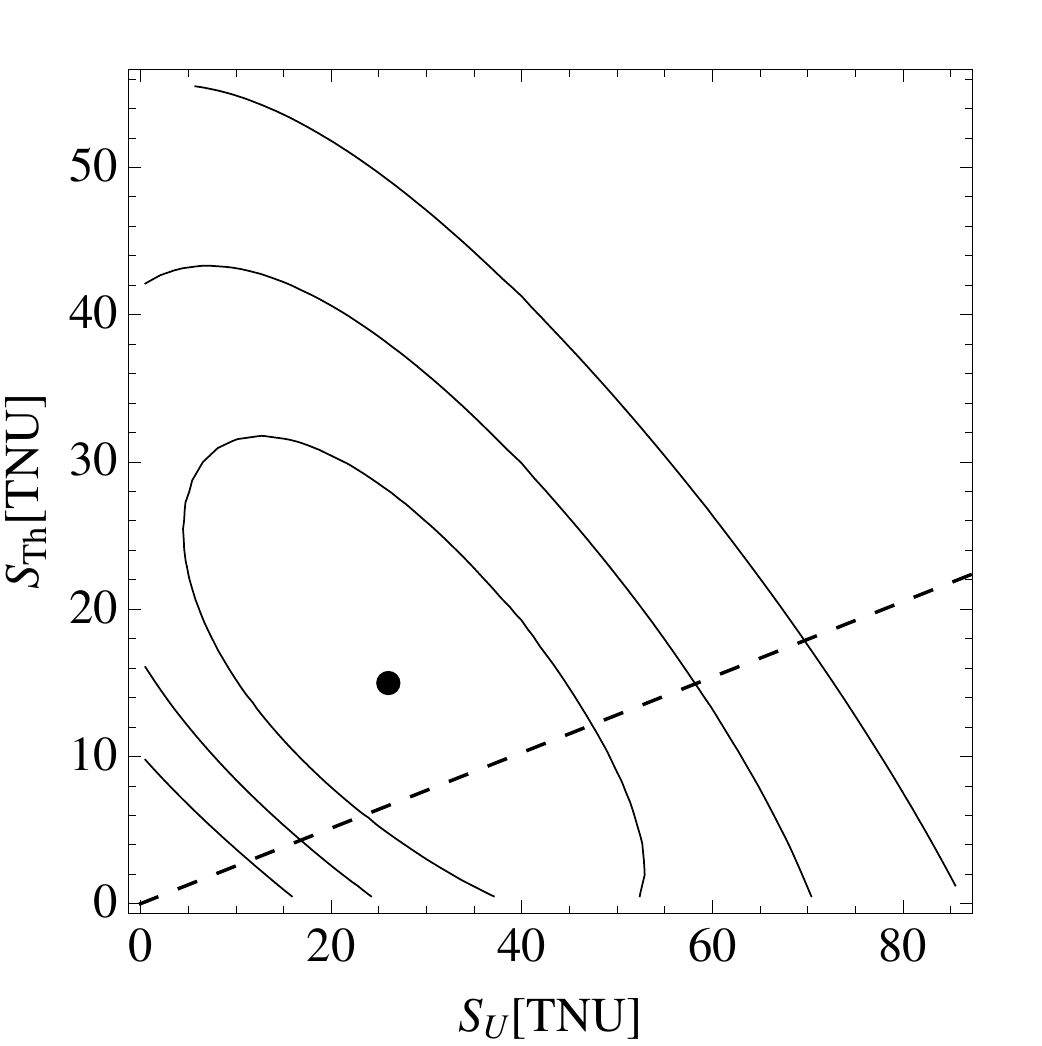}
\caption{Best-fit contours for 1, 2 and 3$\sigma$ for the statistics reported in this paper and for an unbinned likelihood fit with U and Th kept as distinct and free parameters. All other parameters in the fit were kept unchanged. Dashed line corresponds to the chondritic assumption.}
\label{fig3}
\end{figure}

Fig. \ref{fig3} shows the probability contours obtained by performing the fit leaving the U and Th spectral contributions as free parameters. The U and Th best-fit contributions are shown in Fig. \ref{fig1}. This measurement shows how Borexino, with larger exposure, could separate the contributions from U and Th, and demonstrates the ability of this detection technique to perform real-time spectroscopy of geo-neutrinos.

\begin{figure}[!t]
\includegraphics[width=0.5\textwidth]{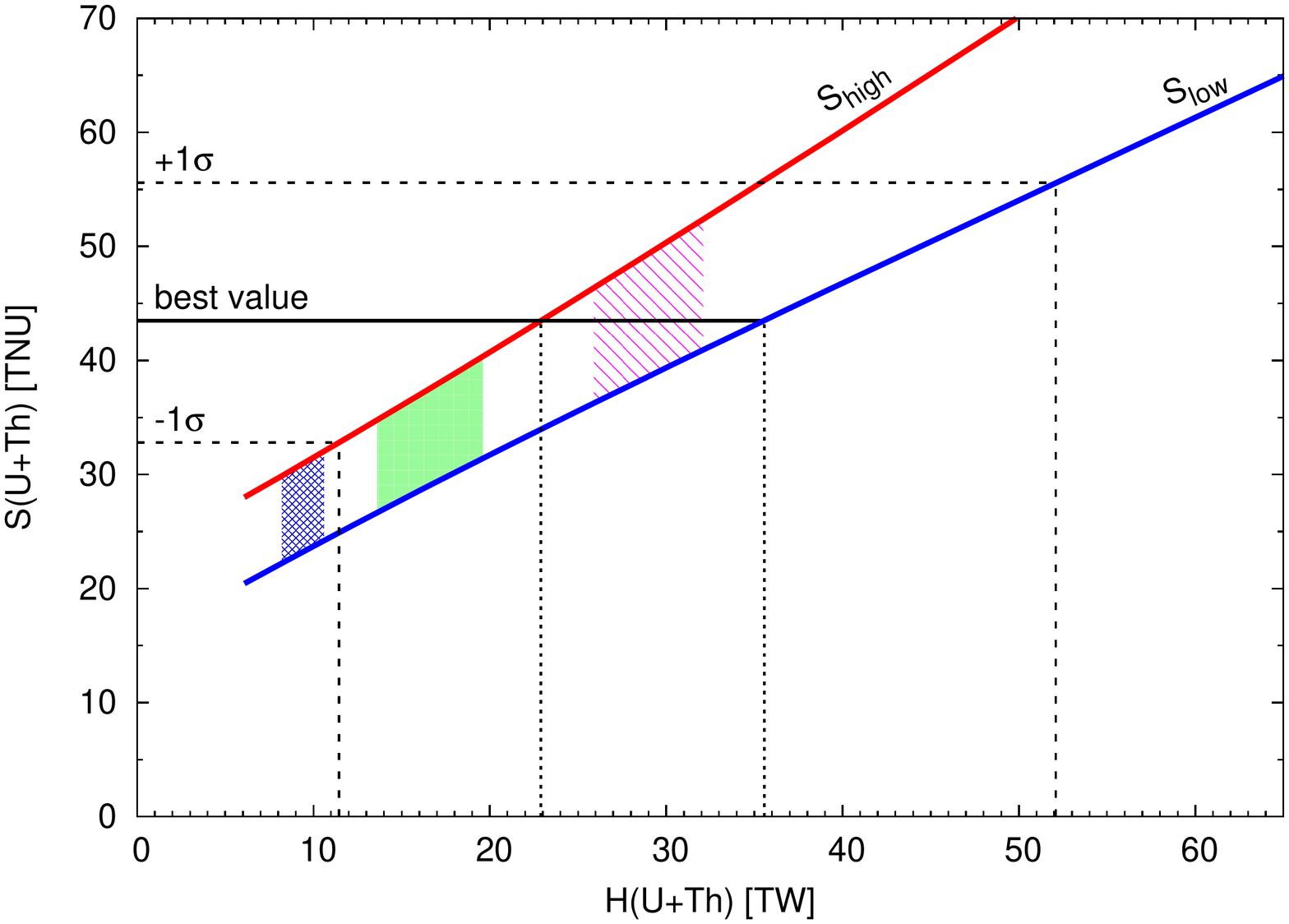}
\caption{The expected geo-neutrino signal in Borexino from U and Th as a function of radiogenic heat released  in radioactive decays of U and Th \cite{georeview2}. The three filled regions delimit, from the left to the right, the cosmochemical, geochemical and geodynamical BSE models \cite{models}. Best values from Borexino together with $\pm 1\sigma$ errors are reported: the experimental statistical and systematic uncertainties have been added in quadrature.} 
\label{fig4}
\end{figure}

The radiogenic heat production for U and Th, $H(U+Th)$, from the present best-fit result is restricted in the range 23-36~TW (see Fig. \ref{fig4}). The range of values includes the uncertainty on the distribution of heat producing elements inside the Earth.
The model-independent analysis yields a radiogenic heat interval 11-52~TW (69\% C.L.) for $H(U+Th)$.
Adopting the chondritic mass ratio above and a potassium-to-uranium mass ratio $m(K)/m(U)=10^4$, the total measured terrestrial radiogenic power is $P(U+Th+K)=33^{+28}_{-20}$~TW, to be compared with the global terrestrial power output $P_{tot}=47\pm2$~TW \cite{terrestrialpower}.

The contribution to the total geo-neutrino signal from the local crust (LOC) is estimated to be $S_{geo}$(LOC) = (9.7$\pm$1.3) TNU  \cite{LOCpaper}. Considering the contribution from the rest of the crust (ROC) \cite{ROCpaper}, the signal from the crust in Borexino is calculated as $S_{geo}$(LOC+ROC) = (23.4$\pm$2.8) TNU. In order to estimate the significance of a positive signal from the mantle we have determined the likelihood of $S_{geo}(Mantle)=S_{geo}-S_{geo}$(LOC+ROC) using the experimental likelihood profile of $S_{geo}$ and a gaussian approximation for the crust contribution. The non-physical region,  $S_{geo}(Mantle)<0$, is excluded. This approach gives a signal from the mantle equal to  $S_{geo}(Mantle) = 20.9^{+15.1}_{-10.3}$ TNU, with the null hypothesis rejected at 98\% C.L..

An updated measurement of  $\bar{\nu}_e$'s with Borexino is presented. We show that Borexino-only data measure geo-neutrinos with 5.9$\sigma$ significance.
We also shows that the background level in Borexino allows to perform a real time spectroscopy of geo-neutrinos, currently limited only by the size of the detector. 

The Borexino program is made possible by funding from INFN (Italy),
NSF (USA), BMBF, DFG, and MPG (Germany), RFBR: Grants 14-22-03031 and 13-02-12140, RFBR-ASPERA-13-02-92440 (Russia), and NCN Poland (UMO-2012/06/M/ST2/00426).  
We acknowledge the financial support from the UnivEarthS Labex program of Sorbonne Paris CitŽ (ANR-10-LABX-0023 and ANR-11-IDEX-0005-02). 
We acknowledge the generous support and hospitality of the Laboratori Nazionali del Gran Sasso (LNGS). We acknowledge the collaboration of J. Mandula for providing detailed data on reactor load factors.

\end{document}